\documentclass[preprint,prd,nofootinbib]{revtex4-1}

\usepackage[colorlinks=true,linkcolor=red]{hyperref}
\usepackage{amsthm,amscd,mathrsfs}
\usepackage{amsfonts,bm,latexsym}
\usepackage{amsmath,amssymb}
\usepackage[mathscr]{euscript}
\usepackage{bm}
\usepackage{graphicx}
\usepackage{subfigure}
\def\bse{\begin{subequations}}
\def\ese{\end{subequations}}
\def\be{\begin{equation}}
\def\ee{\end{equation}}
\def\bea{\begin{eqnarray}}
\def\eea{\end{eqnarray}}

\begin{document}

\title{Entanglement dynamics for circularly accelerated two-level atoms coupled with electromagnetic vacuum fluctuations }

\author{Jiaozhen She, Jiawei Hu\footnote{Corresponding author: jwhu@hunnu.edu.cn}, Hongwei Yu\footnote{Corresponding author: hwyu@hunnu.edu.cn }}

\affiliation{Department of Physics and Synergetic Innovation Center for Quantum Effects and Applications, Hunan Normal University, Changsha, Hunan 410081, China}

\date{\today}

\begin{abstract}

We investigate, in the framework of open quantum systems, the entanglement dynamics of two circularly accelerated two-level atoms with the same centripetal acceleration interacting with a bath of fluctuating electromagnetic fields in the Minkowski vacuum. We assume that the two atoms rotate synchronically with their separation perpendicular to the rotating plane, and study the entanglement degradation, creation, revival, and enhancement by solving the Markovian master equation. In contrast to the scalar-field case, the entanglement dynamics is crucially dependent on the atomic polarizations in the sense that the polarization directions may affect the entanglement decay rate, and may determine the occurrences of entanglement creation, revival and enhancement. Compared with the uniformly accelerated case and the thermal case, the decay rate of entanglement for circularly accelerated atoms is larger, while the revival and enhancement rates are smaller.

\end{abstract}

\maketitle

\section{INTRODUCTION}
Quantum entanglement is one of the most important concepts in quantum theory, and  plays a key role in novel technologies based on quantum effects, such as quantum information and quantum computation \cite{MAquantum information, DBquantum information}. However, quantum entanglement may be ruined due to the unavoidable interaction between the quantum system and its environment, which is one of the main challenges in the realization of quantum information technologies. In particular, it has been found that the entanglement between a pair of atoms can completely disappear within a finite period of time, which is referred to as entanglement sudden death~\cite{TY,JH2007}.
However, entanglement may also be generated between a pair of initially separable atoms  placed in a common bath  via indirect interactions induced by the common bath  \cite{DB2002,Ms2002,SS2002,AM2002,LJ2002,BT2003,FBR2003,ZF2003,RT2004,ZF2008,FR2005,RT2010}. For two atoms in a thermal bath with a nonvanishing separation, entanglement generation only happens in certain circumstances, but entanglement sudden death is a general feature~\cite{RT2010}. Also, the destroyed entanglement may be recreated, known as entanglement revival \cite{ZF2006}.

In the Minkowski vacuum, a uniformly accelerated observer measures a temperature proportional to its acceleration, which is the well-known Unruh effect \cite{Unruh}. Therefore, it is of interest to investigate the entanglement dynamics between accelerating atoms, and compare the results with those in a thermal bath at the Unruh temperature. Recently, Benatti and Floreanini studied the entanglement creation between two uniformly accelerated atoms with vanishing separation coupled with a bath of fluctuating scalar fields in the Minkowski vacuum, and found that the asymptotic entangled state is the same as that in a thermal bath at the Unruh temperature, thus verifying the Unruh effect from a new perspective \cite{bf04}. This work is further extended to the case of two accelerating atoms near a reflecting boundary in Ref. \cite{zhangjialin}, in which it is shown that the conditions for entanglement generation are not exactly the same as those in a thermal bath \cite{zhangjialin2}. The above-mentioned  Refs. \cite{bf04,zhangjialin,zhangjialin2} investigate either the conditions of entanglement generation in the beginning of evolution or the late equilibrium state. Since entanglement generation may not happen at the beginning of evolution but shows a delayed feature, see e.g. Ref.  \cite{ZF2008}, and the entanglement generated during evolution or prepared at the beginning may be destroyed, a systematic investigation of the whole evolution
process is necessary. At this point, we note that the entanglement evolution for accelerating atoms has recently been studied in Refs.~\cite{A.G.S,J.Doukas 2010,D.C.M,JW2015PRA,YQ Y,cheng18}.

Usually, the Unruh effect is concerned with linearly accelerated observers. However, it is also interesting to study the case of circular motion since here a large acceleration,  which is necessary to observe the Unruh effect, is easier to achieve experimentally. The quantization of scalar fields in rotating frames was first investigated by Letaw and Pfautch~\cite{JR1980}. An essential difference between circular and linear accelerations is that the radiation perceived by a circularly orbiting observer is nonthermal \cite{JR1981,SK1987,JS1983,JS1987,WG UNRUH1998,ET2007}, so it is also worth investigating the entanglement dynamics of circularly accelerated atoms. In Ref. \cite{JW2015}, we  studied the dynamics of quantum entanglement and quantum discord of two circularly accelerated two-level atoms coupled with a bath of fluctuating massless scalar fields in the Minkowski vacuum. However, two-level atoms coupled with a bath of fluctuating scalar fields is somewhat a toy model, and a more realistic model of the environment would be a bath of fluctuating vacuum electromagnetic fields. In this paper, we plan to investigate the entanglement dynamics of two mutually independent circularly accelerated two-level atoms rotating synchronically with their separation perpendicular to the rotating plane, and compare the results with those of uniformly accelerated atoms, and static ones immersed in a thermal bath at the Unruh temperature. In particular, we focus on the effect of atomic  polarization on the entanglement dynamics, which is absent in the scalar-field case.

\section{BASIC FORMALISM}

We consider a system with two circularly accelerated atoms weakly coupled to a bath of fluctuating electromagnetic fields in the Minkowski vacuum. The Hamiltonian of the whole system can be written as
\be
H=H_A+H_F+H_I.
\ee
Here $H_A$ denotes the Hamiltonian of the two-atom system,
\be
H_A=\frac{\omega}{2}\sigma_3^{(1)}+\frac{\omega}{2}\sigma_3^{(2)},
\ee
where $\sigma_i^{(1)}
= \sigma_i\otimes \sigma_0,~\sigma_i^{(2)}= \sigma_0\otimes \sigma_i $, with $\sigma_i~ (i = 1,2,3)$ being the Pauli matrices, $\sigma_0 $ being the $2\times2$ unit matrix, and $\omega$ being the energy-level spacing of the atoms. $H_F$ is the Hamiltonian of the electromagnetic fields of which the explicit expression is unnecessary  here. $H_I$ is the dipole  interaction between the atoms and the fluctuating electromagnetic fields, which can be written as
\be
H_I=\mathbf{-D^{(1)}(\tau)\cdot E[x^{(1)}(\tau)]-D^{(2)}(\tau)\cdot E[x^{(2)}(\tau)]},
\ee
where $\mathbf{D^{(\alpha)}(\tau)}~(\alpha=1,2)$ is the electric-dipole moment operator of the $\alpha$th atom, and $\mathbf{E[x^\alpha(\tau)]}$ is the electric-field strength.

For simplicity, we assume that initially the two atoms are decoupled from the quantum electromagnetic fields, $\rho_{\rm tot}(0)=\rho(0)\otimes\rho_f (0)$,  where $\rho(0)$ represents the initial state of the atoms and $\rho_f(0)$ is the Minkowski vacuum state of the electromagnetic fields. The density matrix of the total system satisfies the Liouville equation
\be
\frac{\partial\rho_{\rm tot}(\tau)}{\partial\tau}=-i[H,\rho_{\rm tot}(\tau)].
\ee
Define \be
A^{(\alpha)}(\omega)\equiv A^{(\alpha)}=\mathbf{d^{(\alpha)}}\sigma_- e^{-i\omega\tau},\qquad
A^{(\alpha)}(-\omega)\equiv A^{(\alpha)\dagger}=\mathbf{d^{(\alpha)\ast}}\sigma_+e^{i\omega\tau},
\ee
where $\mathbf{d}^{(\alpha)} =\langle0|\mathbf{D}^{(\alpha)}|1\rangle $ is the transition matrix element of the dipole operator. Under the Born-Markov approximation, the reduced density matrix of the two-atom system $\rho(\tau)=\text{Tr}_F[\rho_{\rm tot}(\tau)]$ satisfies the Kossakowskl-Lindblad master equation \cite{VGA1976,Glc1976,HP2002},
\begin{equation}\label{zhufangchen}
\frac{\partial\rho(\tau)}{\partial\tau}
=-i\big[H_{\rm eff},\,\rho(\tau)\big]+{\cal D}[\rho(\tau)]\,
\end{equation}
where
\be
H_{\rm eff}=H_A-\frac{i}{2}\sum_{\alpha,\beta=1}^2\sum_{i,j=1}^3 H_{ij}^{(\alpha\beta)}\sigma_i^{(\alpha)}\sigma_j^{(\beta)},
\ee
and
\be
\mathcal{D[\rho(\tau)]}=\frac{1}{2}\sum_{\alpha,\beta=1}^2\sum_{i,j}^3 C_{ij}^{(\alpha\beta)}
[2\sigma_j^{(\beta)}\rho\sigma_i^{(\alpha)}-\sigma_i^{(\alpha)}\sigma_j^{(\beta)}\rho-\rho\sigma_i^{(\alpha)}\sigma_j^{(\beta)}].
\ee
Introducing the Fourier transform of the two-point function  $\langle 0| E_{m}(\tau,\mathbf{x}_{\alpha}) E_{n}(\tau',\mathbf{x}_\beta) |0 \rangle$
\be\label{fourier}
\mathcal{G}_{mn}^{(\alpha\beta)}(\omega)=\int_{-\infty}^{\infty}d u\, e^{i\omega u}\langle 0|E_{m}(\tau,\mathbf{x}_{\alpha}) E_{n}(\tau',\mathbf{x}_\beta)|0\rangle,
\ee
where $u=\tau-\tau '$, the coefficient matrix $C^{(\alpha\beta)}_{i j}$ can  be expressed as
\be\label{cxishu}
C^{(\alpha\beta)}_{ij}=A^{(\alpha\beta)}\delta_{ij}-iB^{(\alpha\beta)}\epsilon_{ijk}\delta_{3k}-A^{(\alpha\beta)}\delta_{3i}\delta_{3j},
\ee
where
\begin{align}\label{ABxishu}
A^{(\alpha\beta)}=\frac{1}{4}[\mathcal{G}^{(\alpha\beta)}(\omega)+\mathcal{G}^{(\alpha\beta)}(-\omega)],\quad
B^{(\alpha\beta)}=\frac{1}{4}[\mathcal{G}^{(\alpha\beta)}(\omega)-\mathcal{G}^{(\alpha\beta)}(-\omega)],
\end{align}
with
\be\label{Glhanshu}
\mathcal{G}^{(\alpha\beta)}(\omega)=\sum_{m,n=1}^3 d_m^{(\alpha)\ast} d_n^{(\beta)}\mathcal{G}^{(\alpha\beta)}_{mn}(\omega).
\ee
Similarly, $H^{(\alpha\beta)}_{ij}(\omega)$ can be derived by replacing $\mathcal{G}^{(\alpha\beta)}_{mn}(\omega)$ in Eq. (\ref{Glhanshu}) with
\be
\mathcal{K}^{(\alpha\beta)}_{mn}(\omega)=\frac{P}{\pi i}\int_{-\infty}^\infty d\lambda\frac{\mathcal{G}^{(\alpha\beta)}_{mn}(\lambda)}{\lambda-\omega},
\ee
where $P$ represents the principal value.

\section{ENTANGLEMENT DYNAMICS OF TWO CIRCULARLY ACCELERATED  ATOMS}

We assume that the two atoms rotate synchronically with a separation $L$ perpendicular to the rotating plane, so the trajectories of the two atoms can be described, respectively, as
\bea
&&t_1(\tau)=\gamma \tau,~~x_1(\tau)=R\cos\gamma\tau\Omega ,~~y_1(\tau)=R\sin\gamma \tau \Omega,~~z_1(\tau)=0,\nonumber \\
&&t_2(\tau)=\gamma \tau,~~x_2(\tau)=R\cos\gamma\tau\Omega ,~~y_2(\tau)=R\sin\gamma \tau \Omega,~~z_2(\tau)=L,
\eea
where $R$ is the radius of the circular orbit, $\Omega$ is the angular velocity, and $\gamma=1/\sqrt{1-\Omega^2R^2}$ is the Lorentz factor. In the rest frame of the atom, the centripetal acceleration is $a=\gamma^2\Omega^2 R$.
The correlation functions of electromagnetic fields in the laboratory frame take the form
\bea\label{dcc}
\langle0|E_m(x(\tau))E_n(x(\tau'))|0\rangle
&=&\frac{1}{4\pi^2}(-\partial_0\partial'_0\delta_{mn}+\partial_m\partial'_n) \nonumber \\
&&\quad\times\frac{1}{(x-x')^2+(y-y')^2+(z-z')^2-(t-t'-i\varepsilon)^2}.
\eea
In the following calculations, we need the correlation functions in the proper reference frame of the rotating atoms, which can be obtained with a Lorentz transformation. The explicit results and the procedures for how they are derived are given in  Appendix \ref{app} in detail. Now, we consider the ultrarelativistic limit, i.e., $v\rightarrow1$, since a closed-form computation of the Fourier transformation of the  correlation functions (\ref{liangdianhanshu-1})-(\ref{liangdianhanshu-2}) seems not possible for the generic case. The explicit expressions of the field correlation functions in the ultrarelativistic limit are also given in Appendix \ref{app} since they are rather lengthy, see Eqs. (\ref{2p-limit-1})-(\ref{2p-limit-2}).

In this paper, we assume that the magnitudes of the electric dipoles of the atoms are the same, $\mathbf{d^{(1)}=d^{(2)}=d}$, but the orientations may be different. The coefficients of the dissipator in the master equation  (\ref{zhufangchen}) can then be calculated according to Eqs.~(\ref{cxishu})-(\ref{Glhanshu}),
\begin{align}
C^{(11)}_{ij}=A_1\delta_{ij}-iB_1\epsilon_{ijk}\delta_{3k}-A_1\delta_{3i}\delta_{3j},\\
C^{(22)}_{ij}=A_2\delta_{ij}-iB_2\epsilon_{ijk}\delta_{3k}-A_2\delta_{3i}\delta_{3j},\\
C^{(12)}_{ij}=A_3\delta_{ij}-iB_3\epsilon_{ijk}\delta_{3k}-A_3\delta_{3i}\delta_{3j},\\
C^{(21)}_{ij}=A_4\delta_{ij}-iB_4\epsilon_{ijk}\delta_{3k}-A_4\delta_{3i}\delta_{3j},
\end{align}
where we have defined $A_1=A^{(11)}$, $A_2=A^{(22)}$, $A_3=A^{(12)}$, $A_4=A^{(21)}$, and we have defined $B_1, B_2, B_3$, and $B_4$ similarly for brevity.

To investigate the dynamics of the two-atom system, we work in the coupled basis $\{|G\rangle=|00\rangle,|A\rangle=\frac{1}{\sqrt{2}}(|10\rangle-|01\rangle),|S\rangle=\frac{1}{\sqrt{2}}(|10\rangle+|01\rangle),|E\rangle=|11\rangle\}$, and then a set of equations which are decoupled from other matrix elements can be derived \cite{ZF2002}, see Eqs.  (\ref{rgg})-(\ref{reg}) in Appendix \ref{app2}.

We characterize the degree of entanglement by concurrence \cite{WK1998}, which ranges from $0$ (for separable states) to $1$ for
(maximally entangled states). For the X states, the concurrence takes the form \cite{RT2004}
\bea\label{jiuchanC}
C[\rho(\tau)]=\textrm{max}\{0,K_1(\tau),K_2(\tau)\},
\eea
where
\begin{align}\label{jiuchanK}
K_1(\tau)&=\sqrt{[\rho_{AA}(\tau)-\rho_{SS}(\tau)]^2-[\rho_{AS}(\tau)-\rho_{SA}(\tau)]^2}-2\sqrt{\rho_{GG}(\tau)\rho_{EE}(\tau)},\\
K_2(\tau)&=2|\rho_{GE}(\tau)|-\sqrt{[\rho_{AA}(\tau)+\rho_{SS}(\tau)]^2-[\rho_{AS}(\tau)+\rho_{SA}(\tau)]^2}.
\end{align}

Now we begin our study of the entanglement dynamics for circularly accelerated atoms. In particular, we investigate the phenomena of entanglement degradation, generation, revival and enhancement.

\subsubsection{Entanglement degradation}

First, we consider the entanglement degradation of two-atom systems initially prepared in two kinds of maximally entangled states, i.e., the symmetric state $|S\rangle$ and the antisymmetric state $|A\rangle$.

\begin{figure}
\begin{minipage}[t]{0.5\linewidth}
\centering
\includegraphics[width=3.2in]{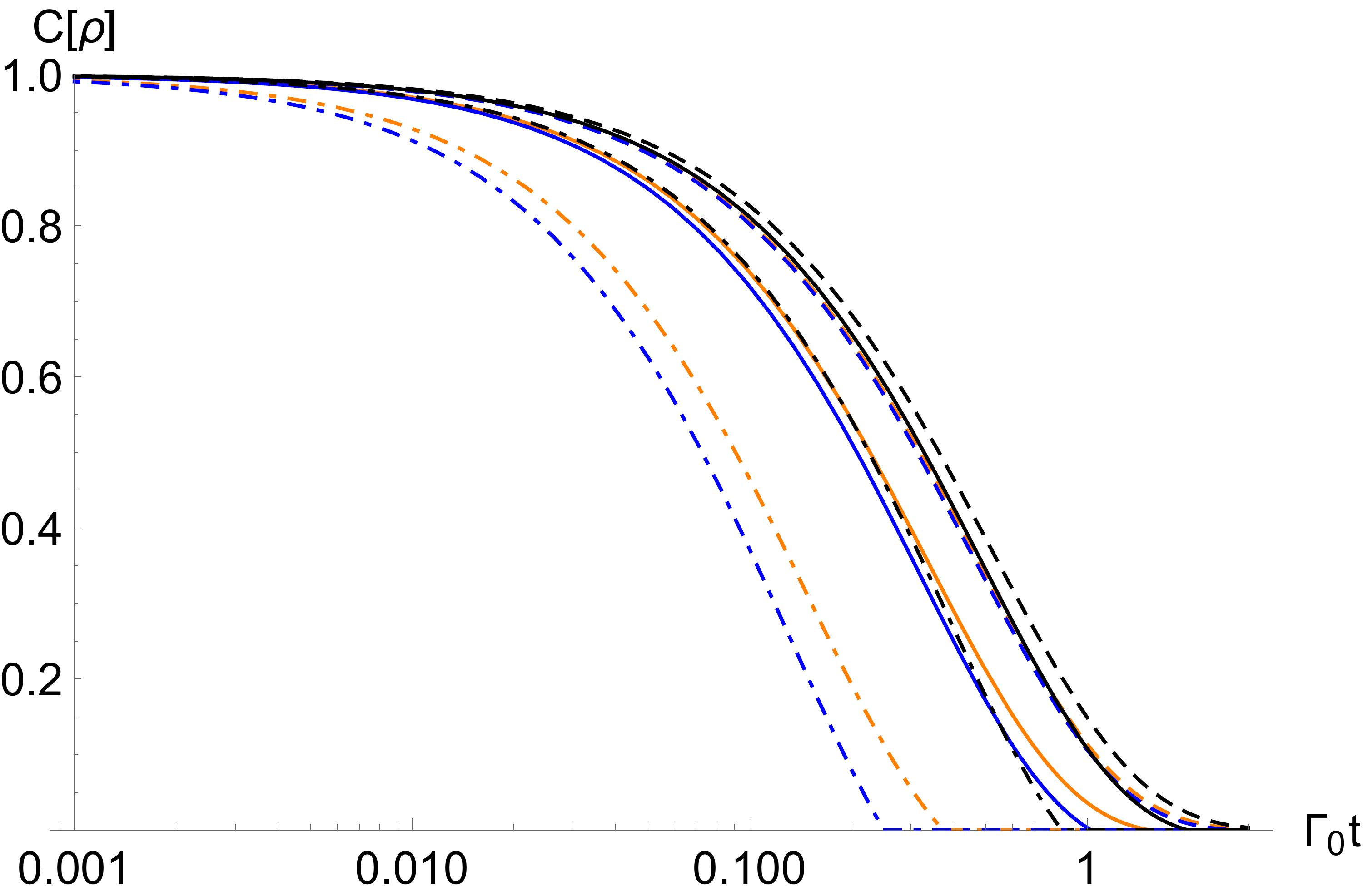}
\end{minipage}%
\begin{minipage}[t]{0.5\linewidth}
\centering
\includegraphics[width=3.2in]{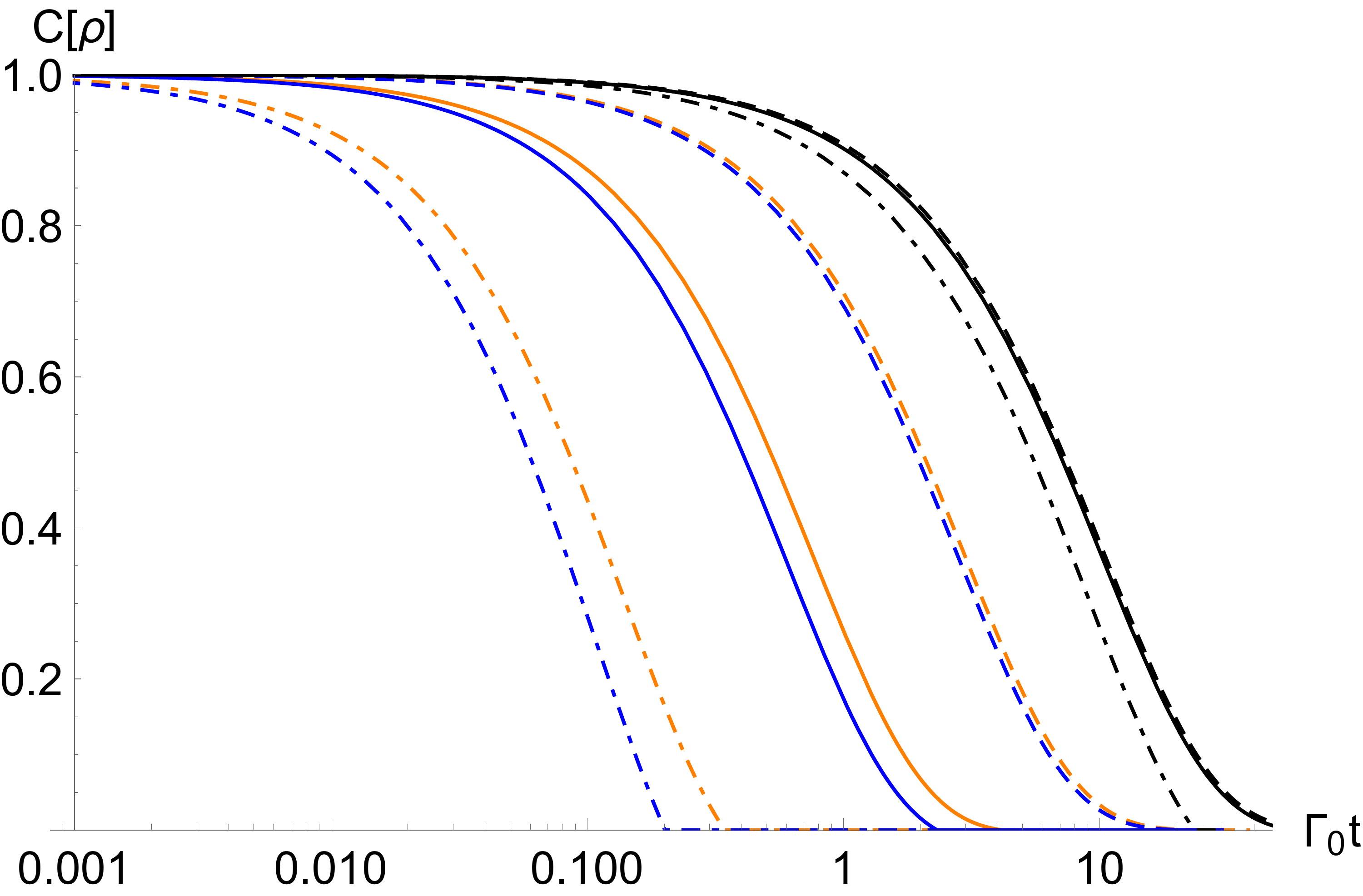}
\end{minipage}
\caption{Comparison between the dynamics of concurrence for circularly accelerated atoms (blue lines), uniformly accelerated atoms (orange lines), and static ones in a thermal bath (black lines) initially prepared in $|S\rangle$ (left) and $|A\rangle$ (right), with $\omega L = 1$. Both of the atoms are polarizable along the $z$ axis. The dashed, solid, and dot-dashed lines correspond to $a/\omega = 1/4, a/\omega = 1$, and $a/\omega = 2$, respectively.}
\label{zs}
\end{figure}

\begin{figure}
\begin{minipage}[t]{0.5\linewidth}
\centering
\includegraphics[width=3.2in]{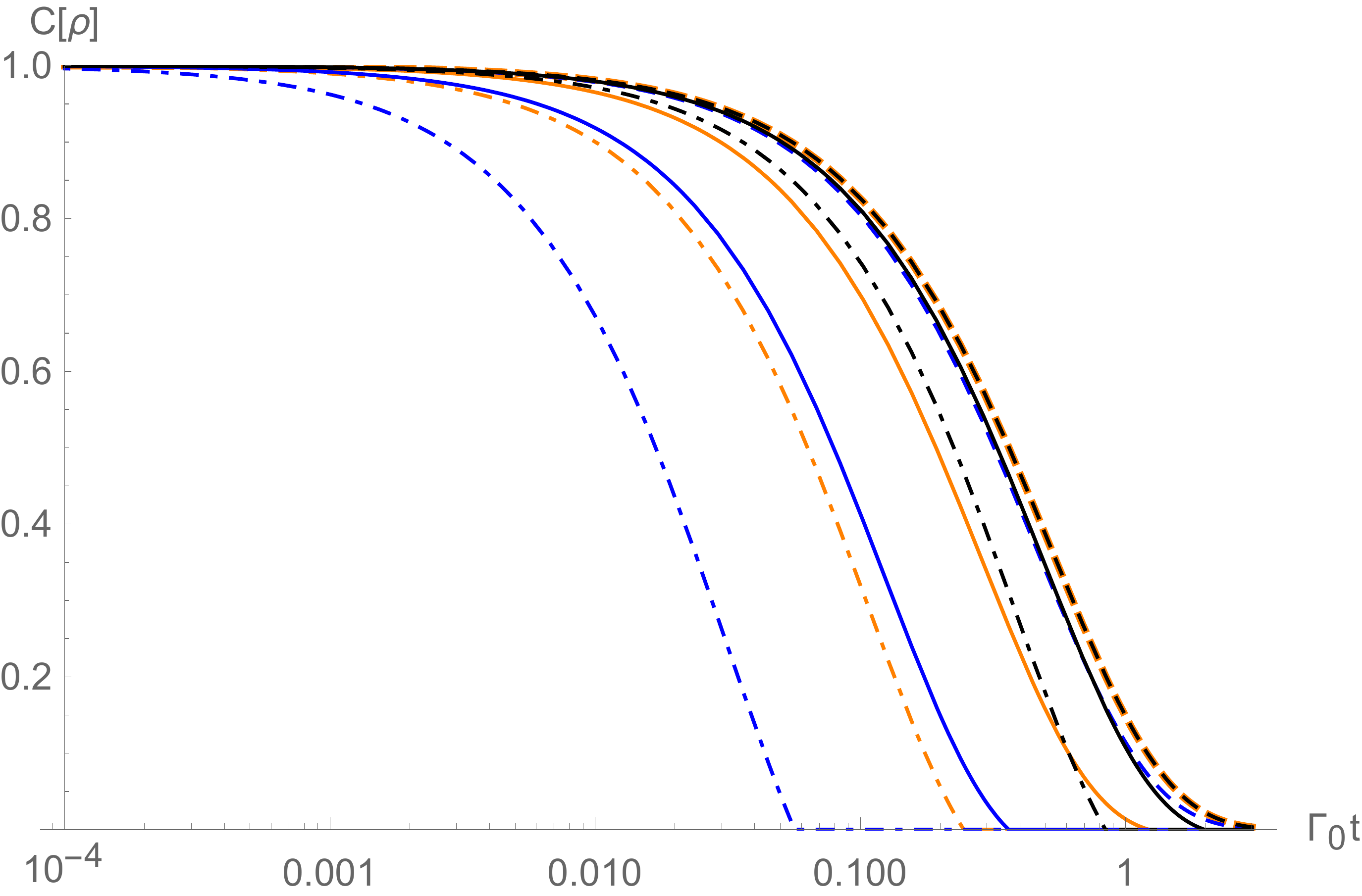}
\end{minipage}%
\begin{minipage}[t]{0.5\linewidth}
\centering
\includegraphics[width=3.2in]{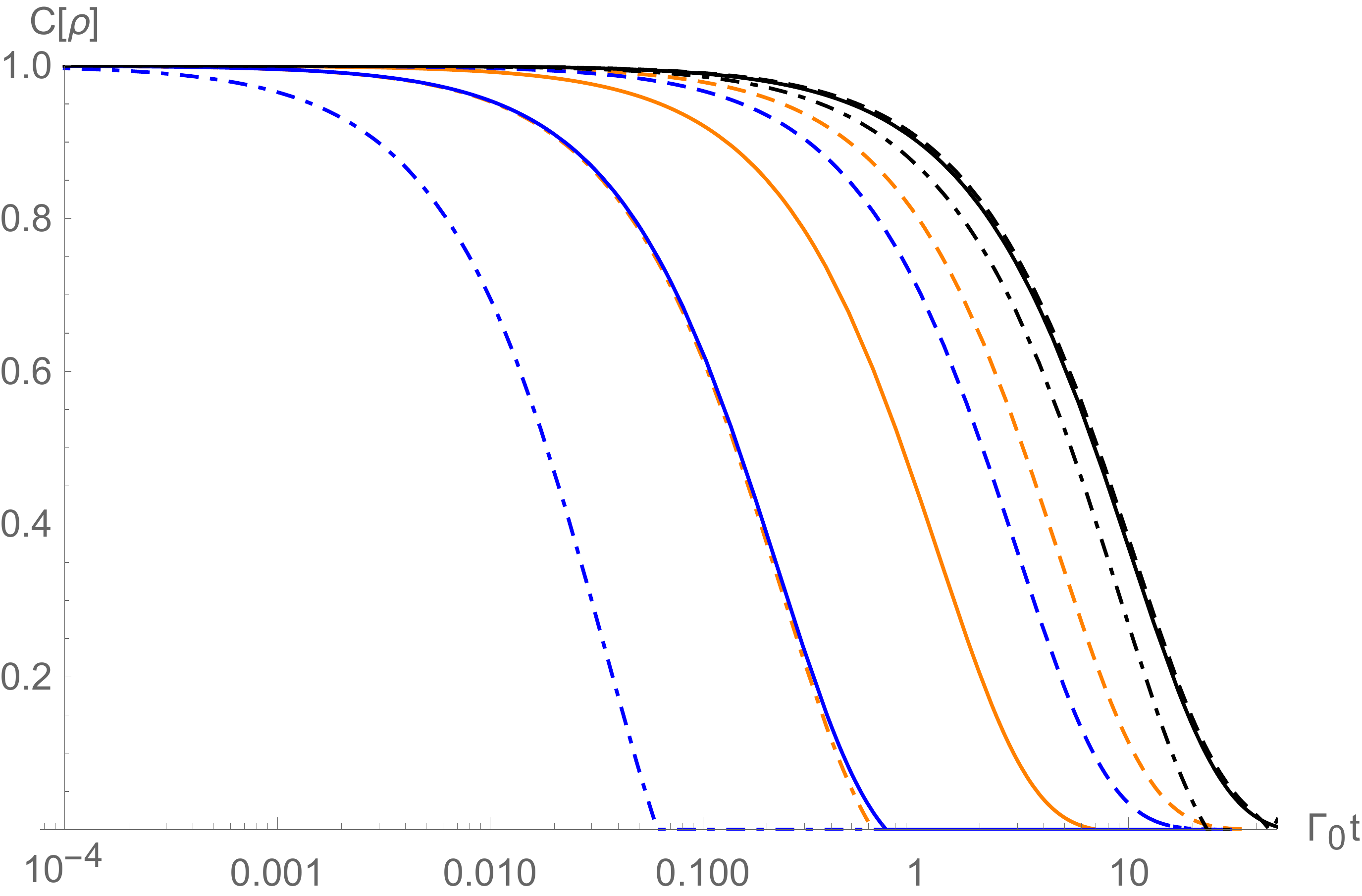}
\end{minipage}
\caption{Comparison between the dynamics of concurrence for circularly accelerated atoms (blue lines), uniformly accelerated atoms (orange lines), and static ones in a thermal bath (black lines) initially prepared in $|S\rangle$ (left) and $|A\rangle$ (right), with $\omega L = 1$. Both of the two atoms are polarizable along the $\varphi$ axis. The dashed, solid, and dot-dashed lines correspond to $a/\omega = 1/4,~a/\omega = 1$, and $a/\omega = 2$, respectively.}
\label{qs}
\end{figure}

When the separation between the two atoms $L$ is very large ($L\rightarrow\infty$), it can be directly obtained from Eqs. (\ref{2p-limit-1})-(\ref{2p-limit-2}) that in this limit, $G^{(12)}_{ij}=G^{(21)}_{ij}=0$,  and thus $A_3,A_4,B_3$, and $B_4$ tend to zero.
Therefore, the evolution of the populations $\rho_{AA}$ (\ref{raa})
and $\rho_{SS}$ (\ref{rss}) are the same, and there is no difference in the
entanglement dynamics whether the initial state is $|A\rangle$ or
$|S\rangle$, which agrees with the linear accelerated case \cite{JW2015PRA,YQ Y}.
For intermediate separations comparable with the transition wavelength ($L\sim \omega^{-1}$), we solve Eqs. (\ref{rgg})-(\ref{reg}) numerically since the analytical solutions are  complicated.
In Figs. \ref{zs} and \ref{qs}, we show that, compared with the results of uniformly accelerated atoms and static ones immersed in a thermal bath at the Unruh temperature $T_U=a/2\pi$, the decay rate of the concurrence of circularly accelerated atoms is faster than those of the accelerated ones and statics ones in a thermal bath, no matter whether the initial state is $|A\rangle$ or $|S\rangle$. Then, we rotate the polarizations of the atoms and find that the decay rates of the concurrence for circularly and uniformly accelerated atoms polarizable along the $\varphi$-axis (Fig. \ref{qs}) are larger than those polarizable along the $z$-axis (Fig. \ref{zs}). When the atoms are polarizable along other directions, the results are essentially the same so they are not shown here. \footnote{As the results of the circular accelerated case and those of the uniformly accelerated and the thermal cases are given in different coordinates, we compare the $\rho$ polarization in the circular accelerated case with the $x$ polarization in the uniformly accelerated case, both of which are the directions of acceleration, and the $\varphi$ polarization with the $y$ polarization, both of which are vertical to the plane defined by  acceleration and the atomic separation. }

\subsubsection{Entanglement generation}

Now, we investigate the entanglement dynamics for two-atom system with the initial state $|E\rangle$, i.e., both of the atoms are initially in the excited state.

\begin{figure}
\begin{minipage}[t]{0.5\linewidth}
\centering
\includegraphics[width=3.2in]{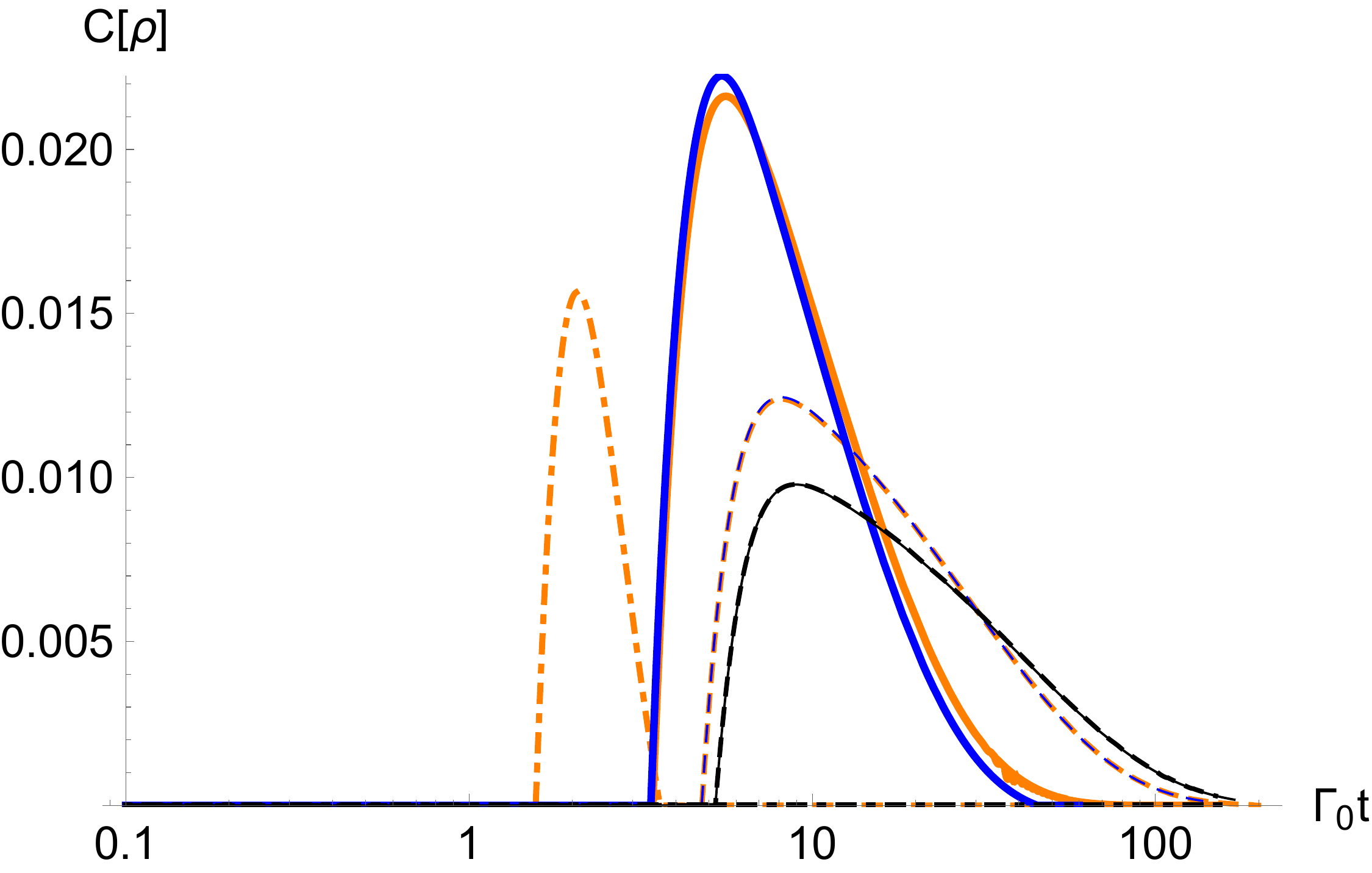}
\end{minipage}%
\begin{minipage}[t]{0.5\linewidth}
\centering
\includegraphics[width=3.2in]{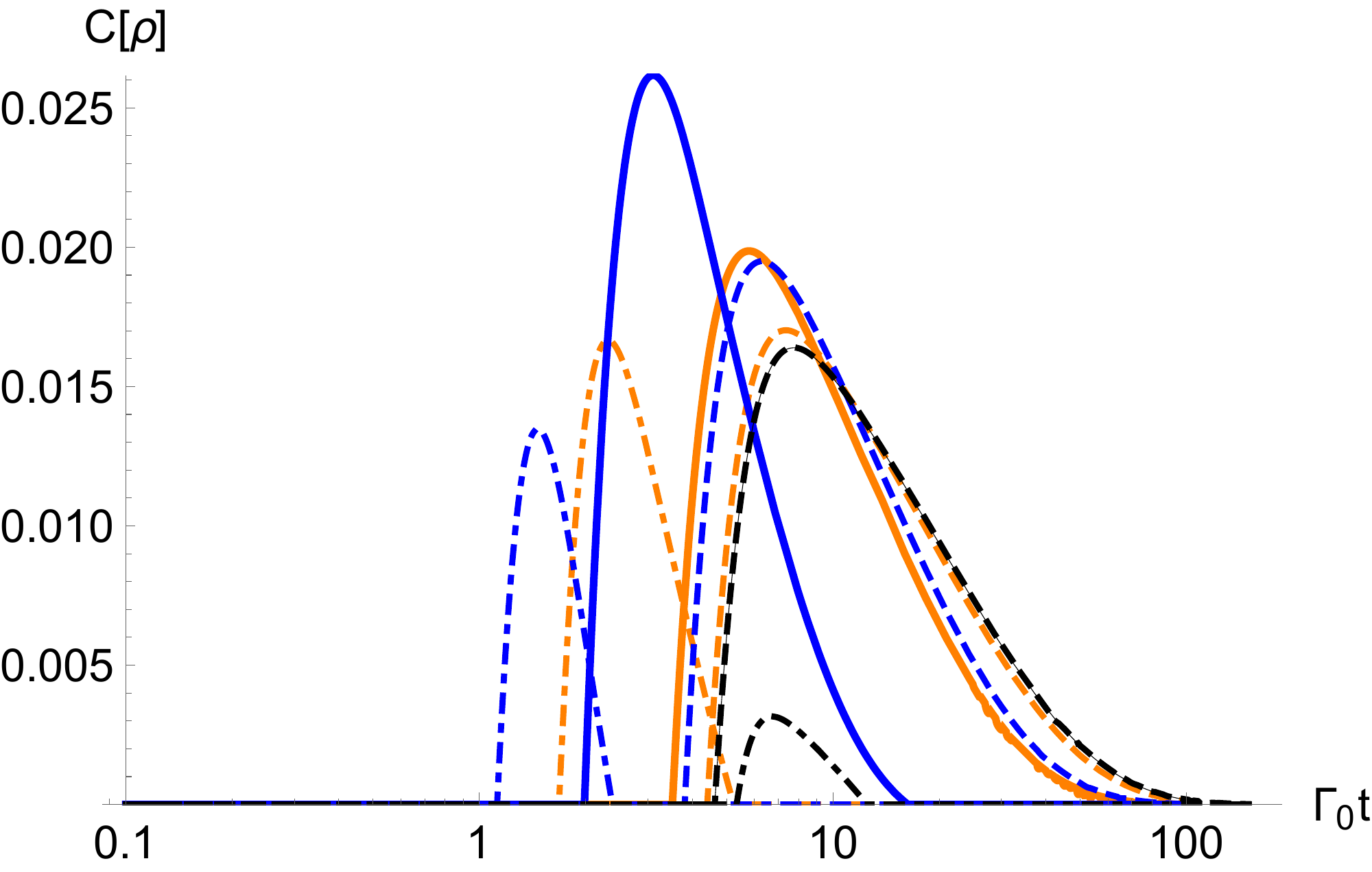}
\end{minipage}
\caption{Comparison between the dynamics of concurrence for circularly accelerated atoms (orange lines), uniformly accelerated atoms (blue lines) and static ones in a thermal bath  (black lines) initially prepared in $|E\rangle$, with $\omega L = 1/2$. Both atoms are polarizable along the $z$-axis (left) or $\varphi$-axis (right). The dashed, solid, and dot-dashed lines correspond to $a/\omega= 1/5, a/\omega = 1/2$, and $a/\omega = 6/5$,
respectively.}
\label{ze}
\end{figure}

As shown in Fig. \ref{ze}, entanglement generation shows a delayed feature, which is known as the delayed sudden birth of entanglement \cite{RT2010}. It is obvious that the lifetime of entanglement of the two-atom system decreases as acceleration increases.  We observe from Fig. \ref{ze}  that entanglement generation depends crucially on the atomic polarization. For circularly accelerated atoms polarizable along the $z$ axis, entanglement generation does not happen when the acceleration is either too small ($a/\omega =1/5$) or too large ($a/\omega =6/5$), while they can get entangled if they are polarizable along the $\varphi$ axis. A comparison between the uniformly accelerated case and the thermal case shows that when the acceleration increases to $a/\omega =6/5$, entanglement generation does not happen for $z$ axis polarizable circularly and uniformly  accelerated atoms, while the static ones in a thermal bath can still get entangled.

\begin{figure}
\begin{minipage}[t]{0.33\linewidth}
\centering
\includegraphics[width=2.18in]{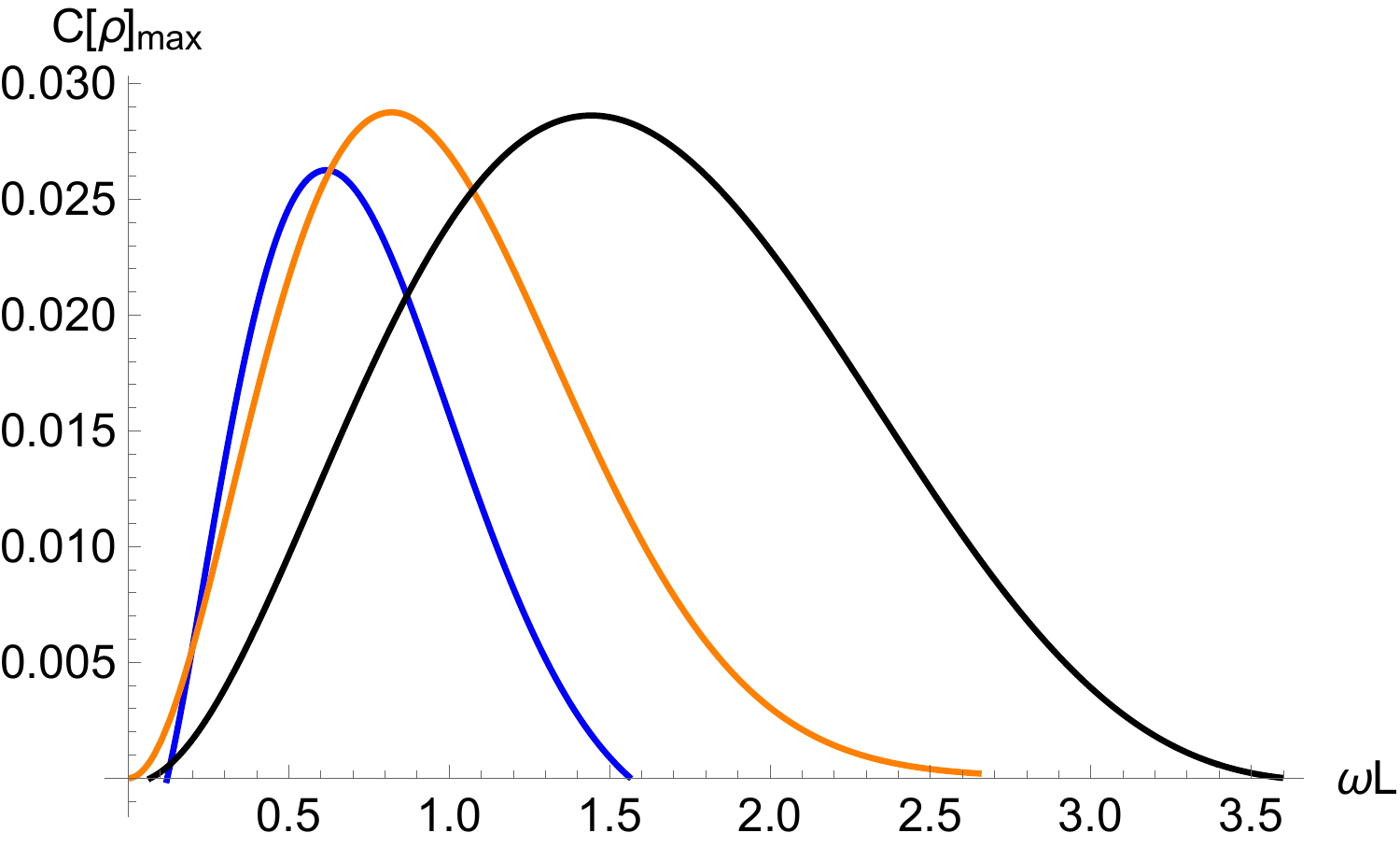}
\end{minipage}%
\begin{minipage}[t]{0.33\linewidth}
\centering
\includegraphics[width=2.18in]{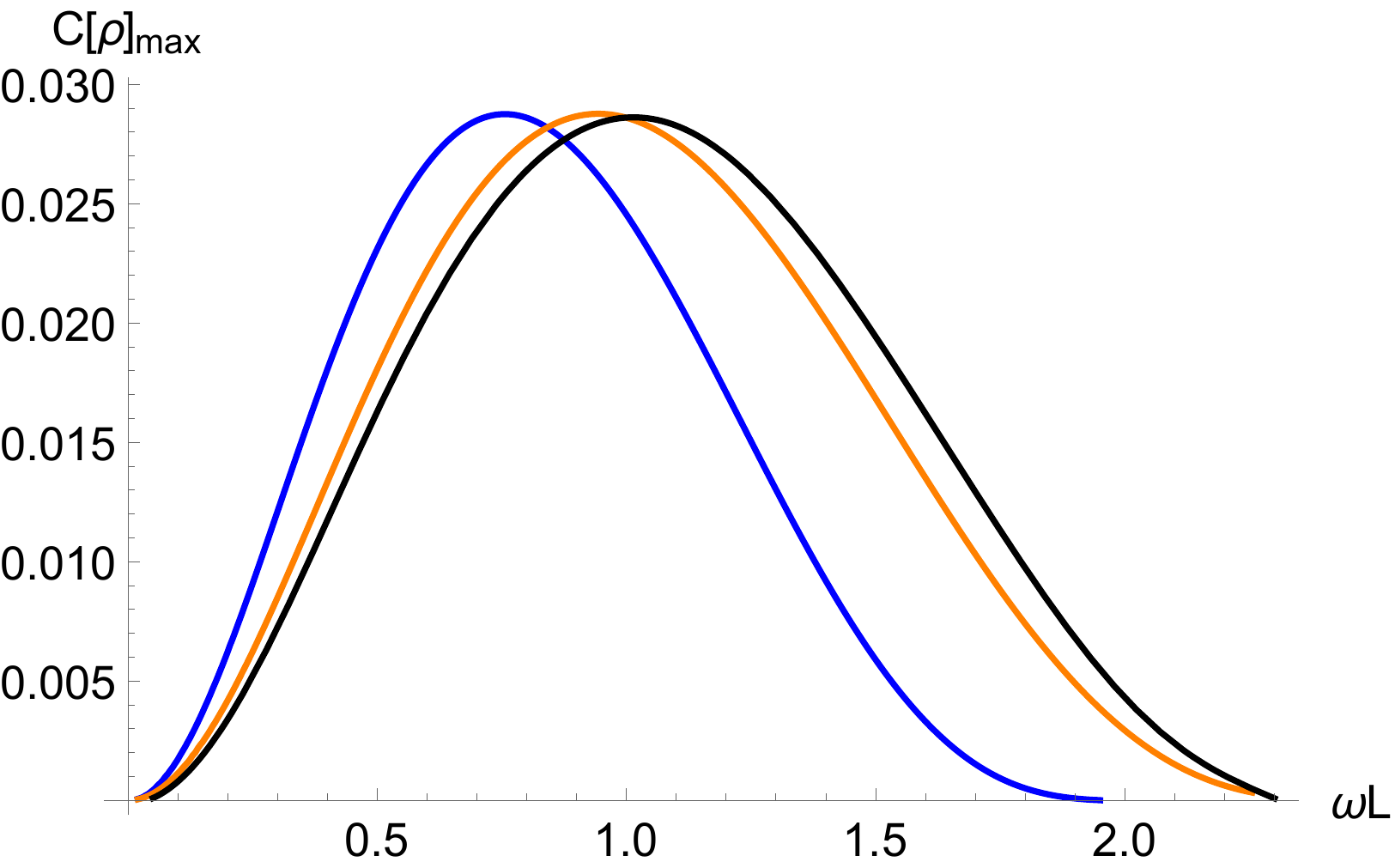}
\end{minipage}
\begin{minipage}[t]{0.33\linewidth}
\centering
\includegraphics[width=2.18in]{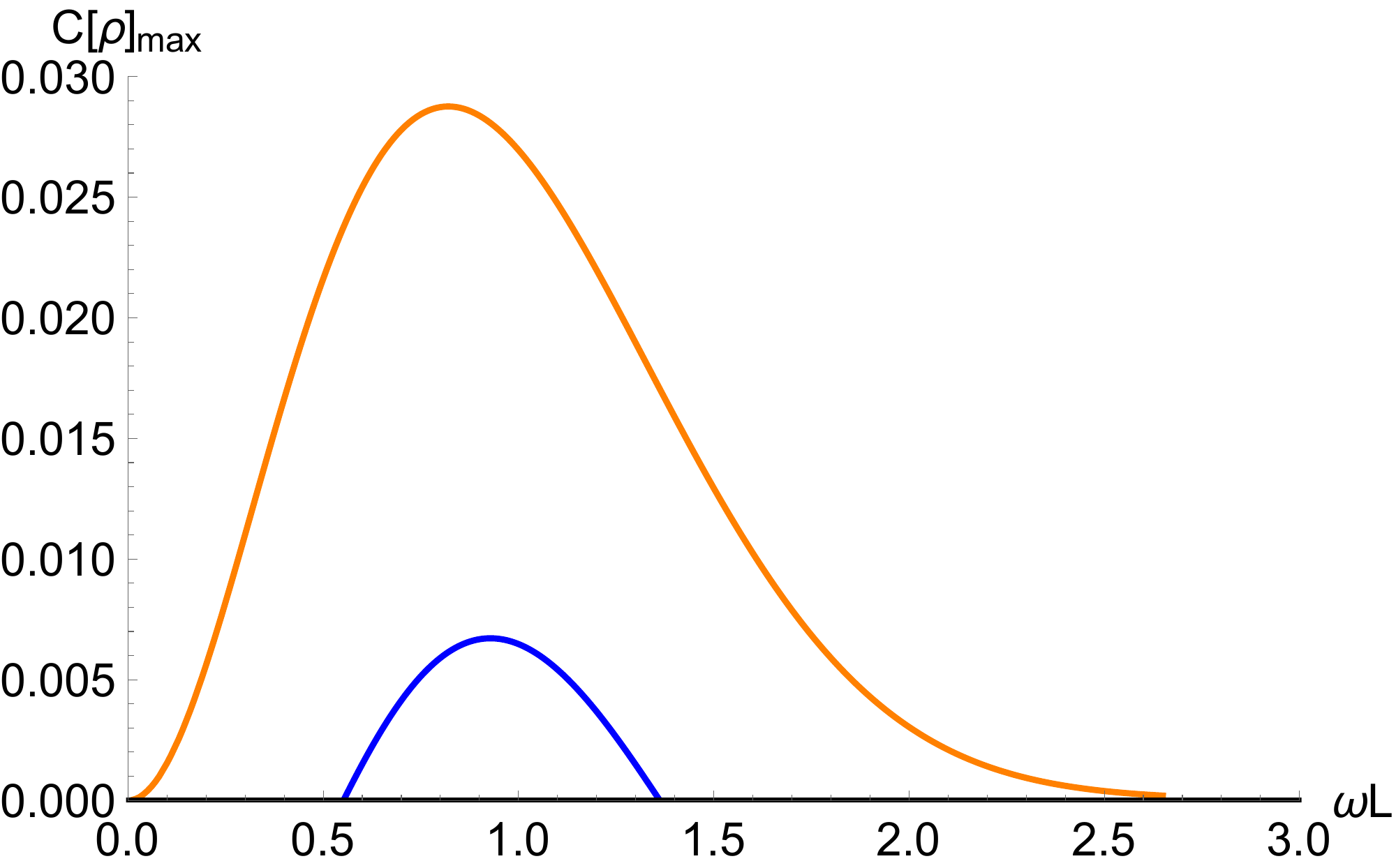}
\end{minipage}
\caption{Comparison between the maximum of concurrence during evolution for circularly accelerated atoms (blue lines), uniformly accelerated atoms (orange lines) and static ones in a thermal bath (black lines)  initially prepared in $|E\rangle$ with $a/\omega = 2/3$. The polarizations of the atoms are $zz$ (left), $\varphi\varphi$ (middle), and $\rho z$ (right) respectively.}
\label{zMaxjiuchan}
\end{figure}

In Fig. \ref{zMaxjiuchan}, we study the effects of atomic separation $L$ on the maximum of entanglement generated during the evolution. The numerical results show that there always exist a minimum and a maximum interatomic separation within which the atoms can be entangled for the circularly accelerated atoms, which agrees with the uniformly accelerated case~\cite{JW2015PRA,YQ Y}. The maximal entanglement during the evolution increases with the separation $L$ first and then reaches its maximum and decreases to zero. When the polarizations of the two atoms are the same, the maximal entanglement for the circularly accelerated atoms is almost the same as those of the uniformly accelerated case  and the thermal case, while the range of $L$ within which entanglement generation happens is smaller (Fig. \ref{zMaxjiuchan}, left and middle). When the polarizations of the two atoms are different, the maximal  entanglement for circularly accelerated atoms is much smaller compared to that of the uniformly accelerated atoms (Fig. \ref{zMaxjiuchan}, right).

\begin{figure}
\begin{minipage}[t]{0.33\linewidth}
\centering
\includegraphics[width=2.18in]{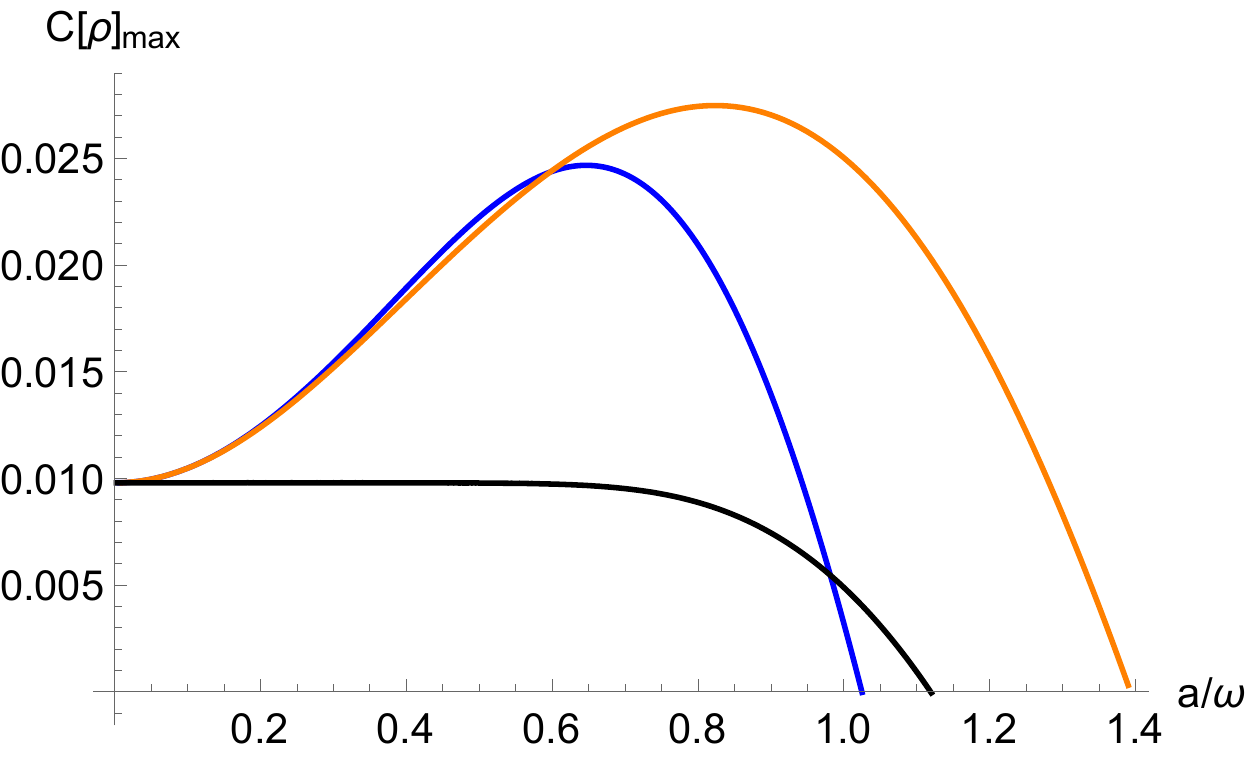}
\end{minipage}%
\begin{minipage}[t]{0.33\linewidth}
\centering
\includegraphics[width=2.18in]{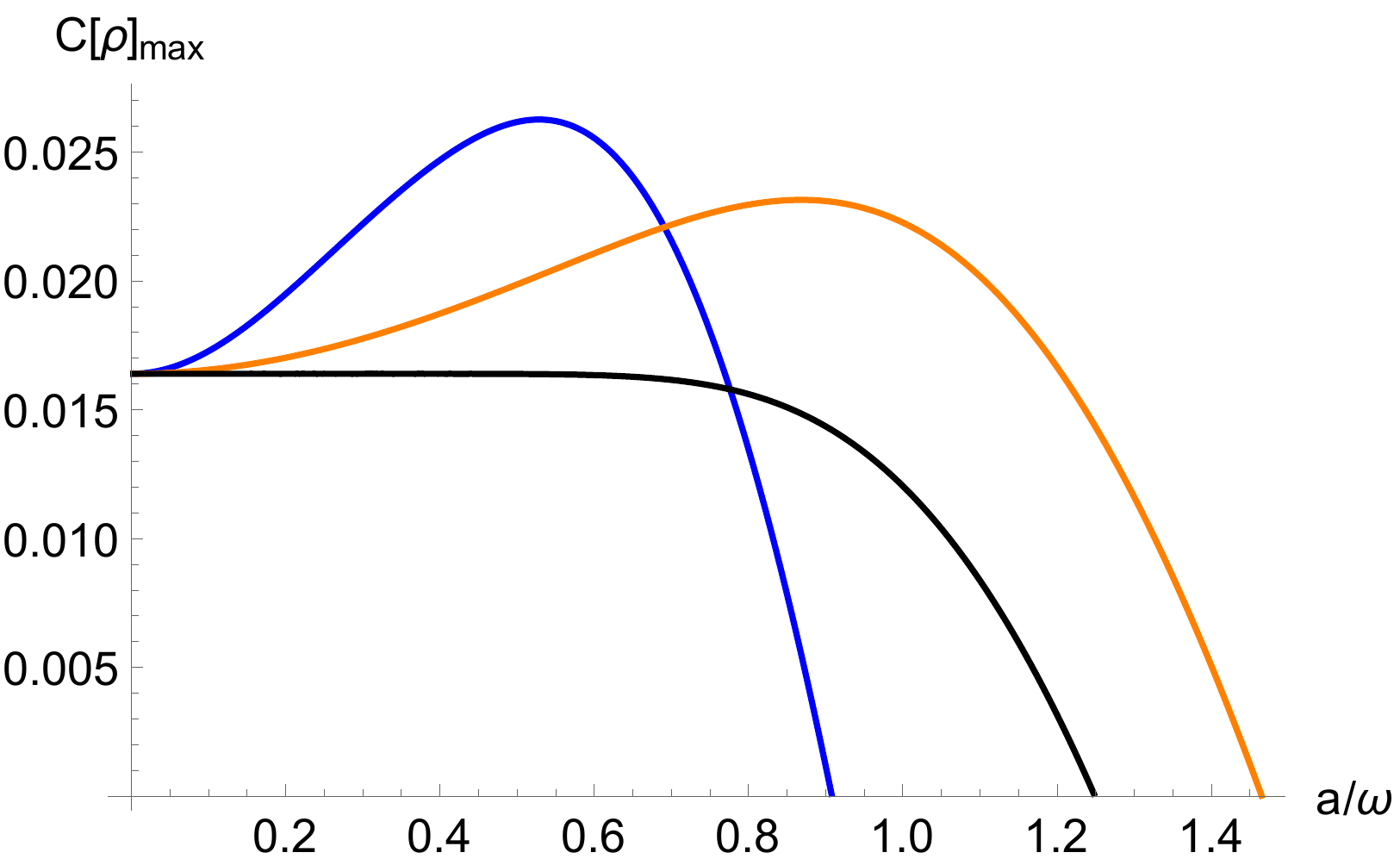}
\end{minipage}
\begin{minipage}[t]{0.33\linewidth}
\centering
\includegraphics[width=2.18in]{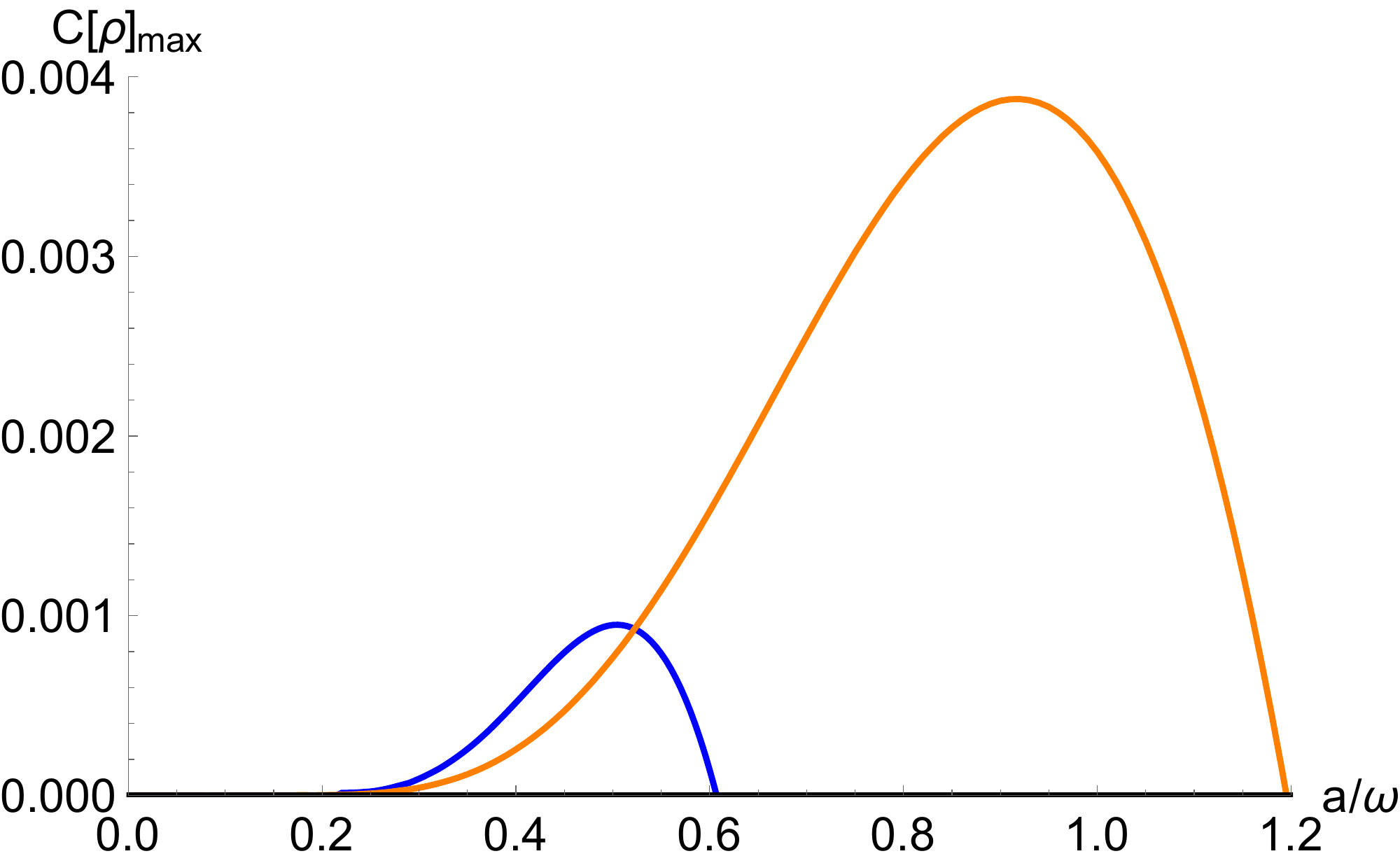}
\end{minipage}
\caption{Comparison between the maximum of concurrence during evolution for circularly accelerated atoms (blue lines), uniformly accelerated atoms (orange lines) and static ones in a thermal bath (black lines) initially prepared in $|E\rangle$ with $\omega L= 1/2$. The polarizations of the atoms are $zz$ (left), $\varphi\varphi$ (middle), and $\rho z$ (right) respectively. }
\label{qMaxjiuchan}
\end{figure}

In Fig. \ref{qMaxjiuchan}, we study the effects of acceleration $a$ on the maximum of entanglement generated during the evolution. Similar to the uniformly accelerated case \cite{YQ Y,JW2015PRA}, the relation between the maximum of the concurrence and acceleration for circularly accelerated atoms is not a monotonically decreasing function as in the thermal case. Also, it is worth to note that, when one of the atoms is polarizable along the direction of the centripetal acceleration and the other is polarizable along the atomic separation, entanglement can be generated for both circularly and uniformly accelerated atoms, but it cannot happen for static atoms in a thermal bath if the two atoms are polarizable along different directions.

\subsubsection{Entanglement revival and enhancement}

\begin{figure}
\begin{minipage}[t]{0.5\linewidth}
\centering
\includegraphics[width=3.2in]{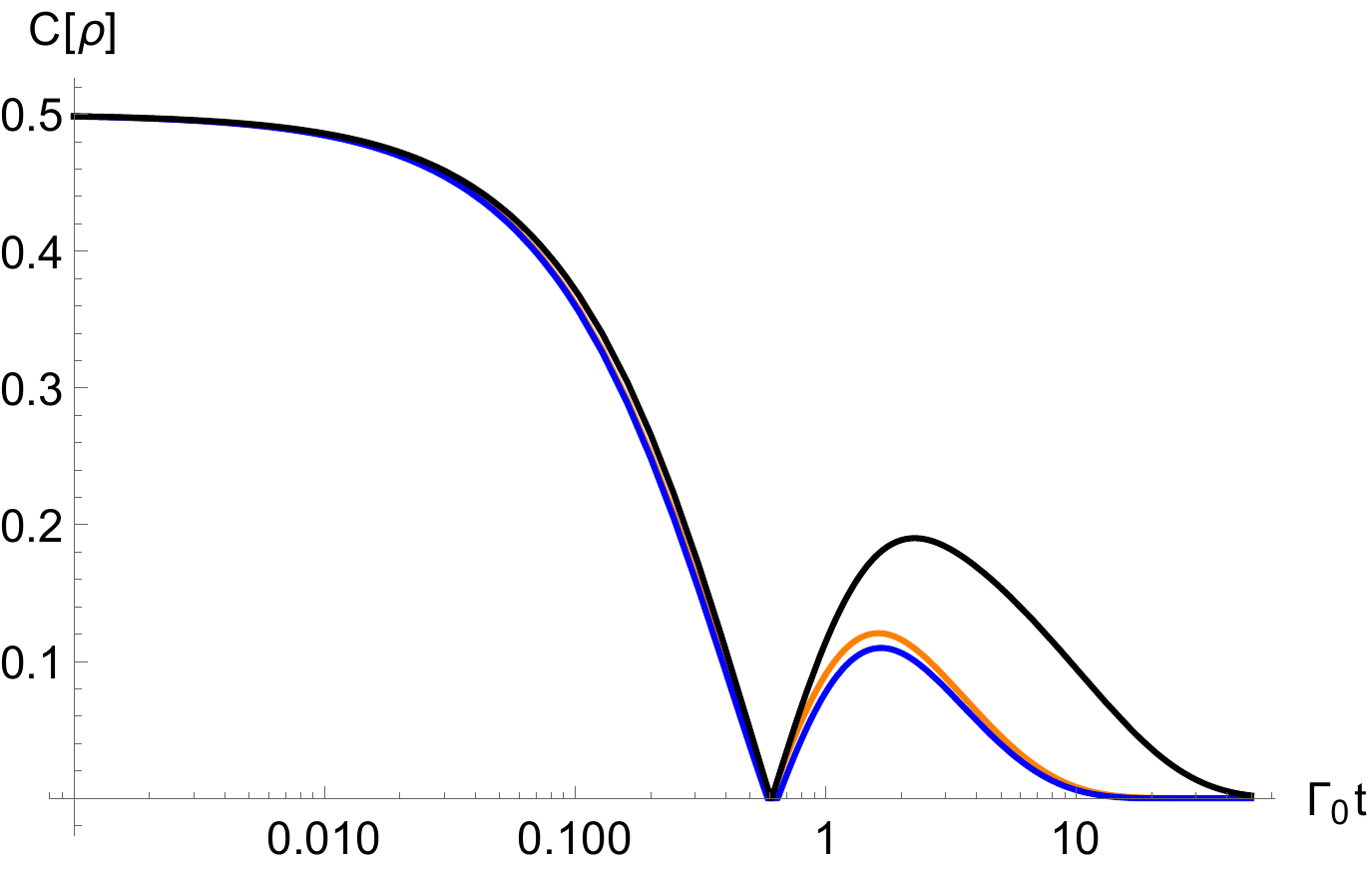}
\end{minipage}%
\begin{minipage}[t]{0.5\linewidth}
\centering
\includegraphics[width=3.2in]{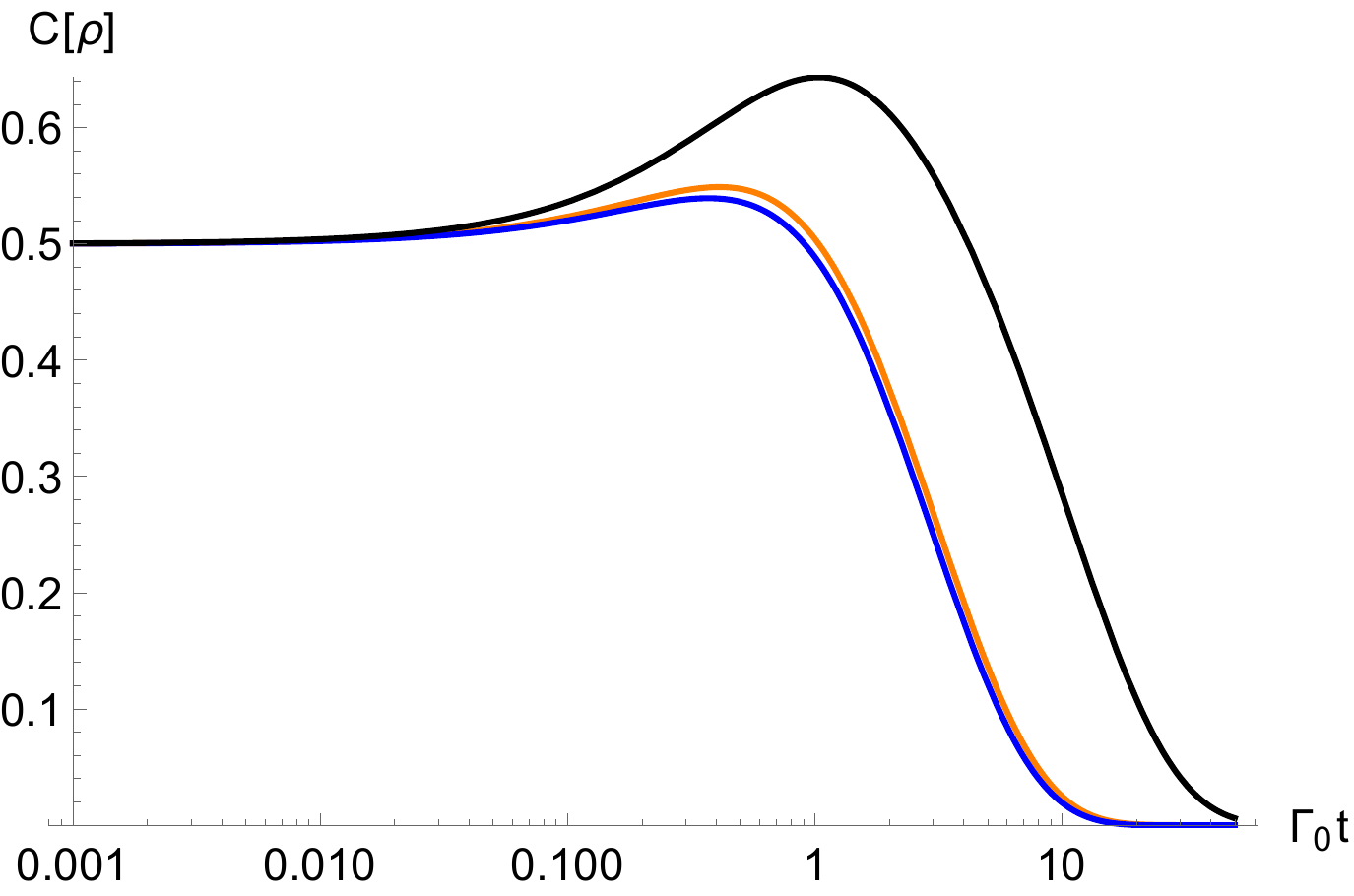}
\end{minipage}
\caption{Comparison between the dynamics of concurrence for circularly accelerated atoms (blue lines), uniformly accelerated atoms (orange lines), and static ones in a thermal bath (black lines) initially prepared in $\frac{1}{2}|A\rangle+\frac{\sqrt{3}}{2}|S\rangle$ (left) and $\frac{\sqrt{3}}{2}|A\rangle+\frac{1}{2}|S\rangle$ (right), with $a/\omega = 1/2$ and $\omega L = 1$. Both atoms are polarizable along the $z$ axis.}
\label{zdiejia13}
\end{figure}

\begin{figure}
\begin{minipage}[t]{0.5\linewidth}
\centering
\includegraphics[width=3.2in]{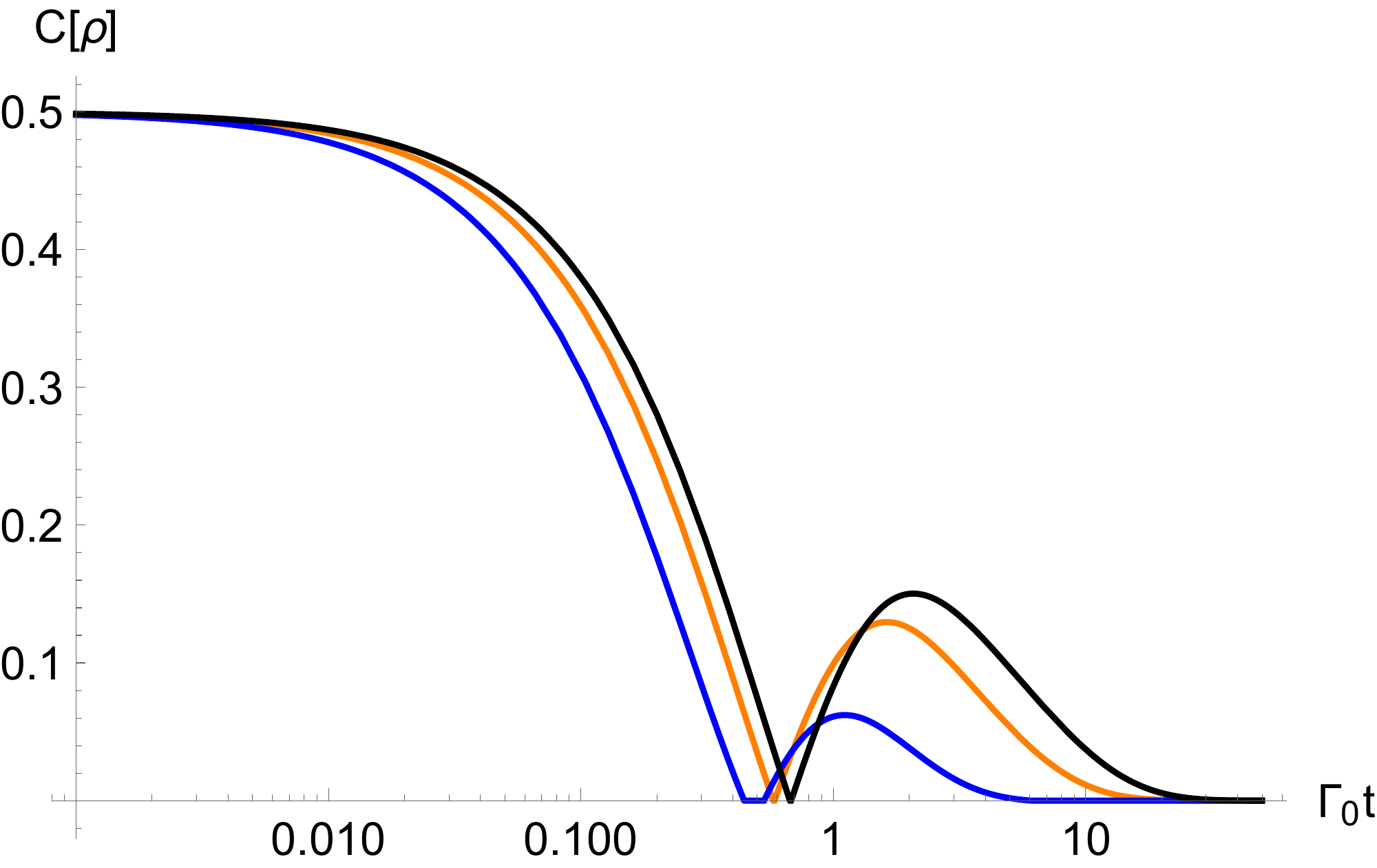}
\end{minipage}%
\begin{minipage}[t]{0.5\linewidth}
\centering
\includegraphics[width=3.2in]{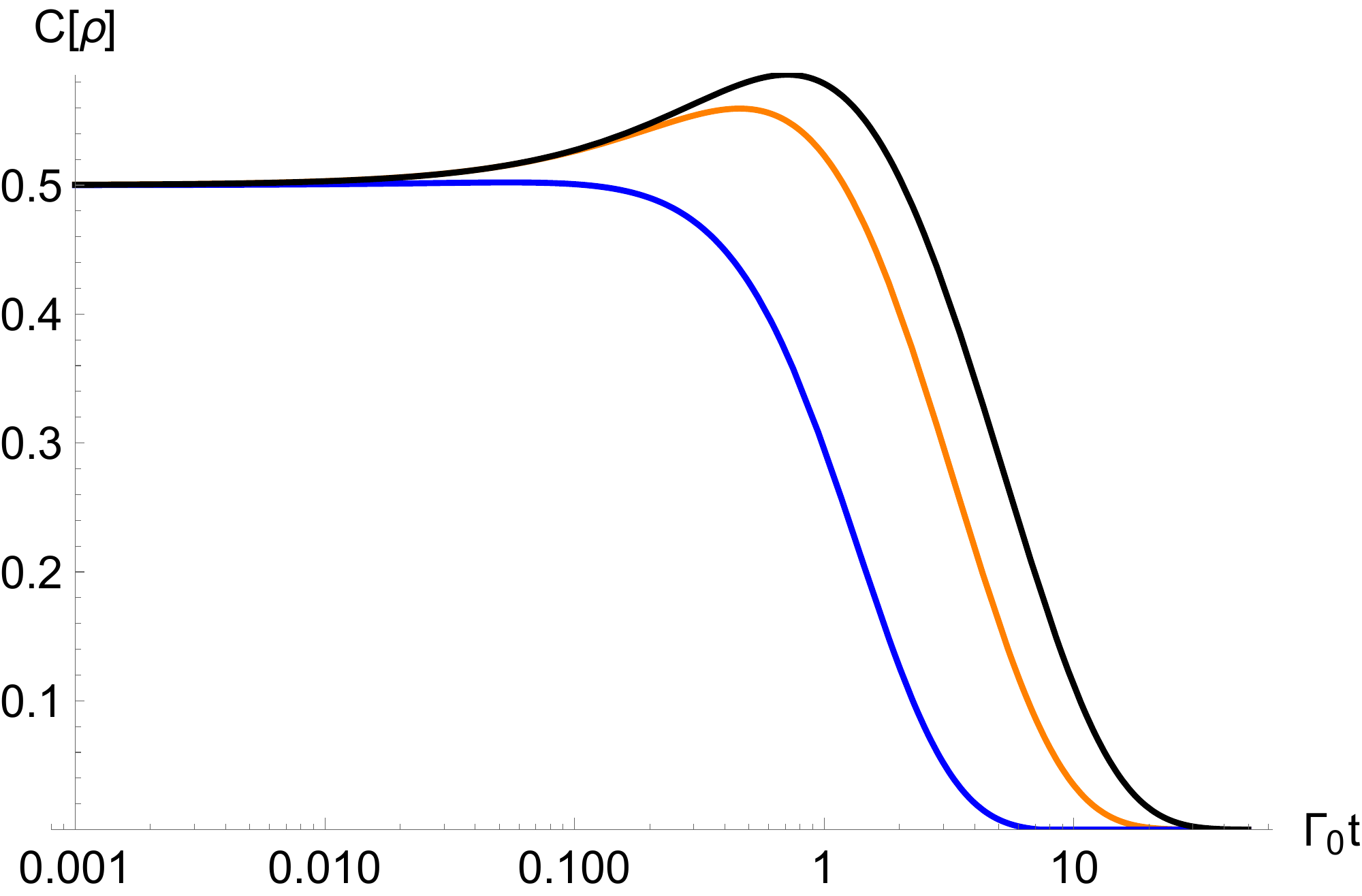}
\end{minipage}
\caption{Comparison between the dynamics of concurrence for circularly accelerated atoms (blue lines), uniformly accelerated atoms  (orange lines), and static ones in a thermal bath (black lines) initially prepared in $\frac{1}{2}|A\rangle+\frac{\sqrt{3}}{2}|S\rangle$  (left) and  $\frac{\sqrt{3}}{2}|A\rangle+\frac{1}{2}|S\rangle$ (right), with $a/\omega = 1/2$ and $\omega L = 1$. Both atoms are polarizable along the $\varphi$ axis.}
\label{qdiejia13}
\end{figure}

In Figs. \ref{zdiejia13} and \ref{qdiejia13}, we investigate the entanglement revival and enhancement for atoms initially prepared in the following states,
\begin{align}
&|\psi\rangle=\sqrt{p}|A\rangle+\sqrt{1-p}|S\rangle~~~(0<p<1,p\neq1/2),
\end{align}
which are entangled states.
When both atoms are polarizable along the $z$ axis (Fig. \ref{zdiejia13}), for $p = 1/4$, the concurrence of the two-atom system first deceases and then revives for a finite period of time. The decay rate of entanglement for circularly accelerated atoms is larger, while the revival rate is smaller, compared with those of uniformly accelerated atoms as well as static ones immersed in a thermal bath. For $p = 3/4$, the initial entanglement can be enhanced during evolution. Compared with the uniformly accelerated case and the thermal case, the enhancement rate of entanglement and the maximum entanglement during evolution for circularly accelerated atoms are smaller. When the atoms are polarizable along the tangential direction $\varphi$, the results are shown in Fig. \ref{qdiejia13}. For $p = 1/4$, the revival of entanglement is apparently  delayed for circularly accelerated atoms. When $p = 3/4$, entanglement enhancement does not happen for circularly accelerated atoms, in contrast to the uniformly accelerated case and the thermal case.


\section{CONCLUSION}

To summarize, in this paper, we have investigated the entanglement dynamics of two circularly accelerated two-level atoms coupled with a bath of fluctuating electromagnetic fields in the Minkowski vacuum in the framework of open quantum systems. For atoms initially in an entangled state, entanglement death within a finite period of time is a general feature for circularly accelerated atoms. The decay rate of entanglement is dependent on the initial state, acceleration, atomic separation, and polarization directions of the atoms. For atoms initially in a separable state $|E\rangle$, entanglement can be generated if the values of acceleration and atomic separation are appropriately assigned. The atomic polarization directions play an important role in both the lifetime of entanglement and the maximal entanglement generated during evolution. When the atoms are initially prepared in  $|A\rangle$ and $|S\rangle$, the phenomena of entanglement revival and enhancement depend crucially on the atomic polarizations. Compared with the uniformly accelerated case and the thermal case, the decay rate of entanglement for circularly accelerated atoms is larger, while the revival and enhancement rates are smaller,

\begin{acknowledgments}

This work was supported in part by the NSFC under Grants No. 11435006, No. 11690034, and No. 11805063.

\end{acknowledgments}

\appendix

\section{ Derivation of the two-point functions in the comoving frame of the rotating atoms}\label{app}

The electromagnetic-field strength $E_i$ is the $i0$ component of the electromagnetic  tensor $F_{\mu\nu}$ defined with the vector potential $A_\mu$ as $F_{\mu\nu}=\partial_\mu A_\nu-\partial_\nu A_\mu$. In matrix form, the electromagnetic tensor $F_{\mu\nu}$ can be explicitly written as
\be
F_{\mu\nu}=
\left(
  \begin{array}{cccc}
     0           & -E_1        & -E_2     & -E_3\\
     E_1         & 0           & B_3      & -B_2\\
     E_2         & -B_3        & 0        & B_1\\
     E_3         & B_2         & -B_1     & 0
  \end{array}
\right).
\ee

After a Lorentz transformation, the electromagnetic tensor in the comoving frame of the rotating atom $F'_{\mu\nu}$ takes the form
\be\label{Au}
F'_{\iota\kappa}=\Lambda_\iota^\mu\Lambda_\kappa^\nu F_{\mu\nu} ,
\ee
where $F_{\mu\nu}$ is the electromagnetic tensor in the laboratory frame, and 
$\Lambda_\iota^\mu$ is the boost matrix
\be\label{Ax}
\Lambda_\iota^\mu=\left(
  \begin{array}{cccc}
    \gamma           & -\gamma\beta n_x  & -\gamma\beta n_y  & 0\\
    -\gamma\beta n_x & 1+(\gamma-1)n_x^2 & (\gamma-1)n_x n_y & 0\\
    -\gamma\beta n_y & (\gamma-1)n_y n_x & 1+(\gamma-1)n_y^2 & 0\\
    0                &         0         &       0           & 1
  \end{array}
\right).
\ee
Here $n_x=v_x/v=-\sin(\Omega\gamma\tau)$, $n_y=v_y/v=\cos(\Omega\gamma \tau)$ are the unit vector of the velocity,  $\gamma=1/\sqrt{1-\beta^2}$ is the Lorentz factor, and in the natural units $\beta=v$. The Lorentz transformation in an arbitrary direction can be found, e.g., in Ref. \cite{MTW}.

Then we transform from the Cartesian coordinate to the cylindrical coordinate. The electric field $E''_i$ ($i=\rho,\varphi,z$) in the frame of the atom expressed in the cylindrical coordinate can be written as $E''_i=S_{i}^jE'_j$, where 
\begin{equation}
S_{i}^j=\left(
  \begin{array}{ccc}
    \cos(\Omega\gamma\tau) & -\sin(\Omega\gamma\tau) & 0\\
    \sin(\Omega\gamma\tau) &  \cos(\Omega\gamma\tau) & 0\\
       0                   &       0                 & 1
  \end{array}
\right)
\end{equation}
is the rotation matrix, and $E'_i$ ($i=x,y,z$) is the electric field  in the Cartesian coordinate. Therefore, one obtains 
\begin{eqnarray}
  &&\langle 0| E''_i(x(\tau)) E''_k(x(\tau')) |0\rangle
=\langle 0| S_{i}^j(\tau)E'_j(x(\tau))\, S_{k}^m(\tau')E'_m(x(\tau')) |0\rangle\nonumber\\
&&\quad=\langle 0| S_{i}^j(\tau)\Lambda_j^\mu(\tau)\Lambda_0^\nu(\tau) F_{\mu\nu}(x(\tau))\, S_{k}^m(\tau')\Lambda_m^\alpha(\tau')\Lambda_0^\epsilon(\tau') F_{\alpha\epsilon} (x(\tau')) |0\rangle.
\end{eqnarray}
That is, the electric-field two-point functions in the frame of the rotating atoms can be related to the two-point function for the vector potential $\langle0| A_\mu(x(\tau)) A_{\nu}(x(\tau'))|0\rangle$ in the laboratory frame, the explicit form of which is
\begin{equation}
\langle0| A_\mu(x(\tau)) A_{\nu}(x(\tau'))|0\rangle
= \frac{1}{4\pi^2} \frac{\eta_{\mu\nu}}{(x-x')^2+(y-y')^2+(z-z')^2-(t-t'-i\varepsilon)^2},
\end{equation}
where $\eta_{\mu\nu}={\rm diag}(-1,+1,+1,+1)$. Finally, we obtain the two-point functions in the comoving frame of the rotating atoms $G^{(\alpha\beta)}_{ij}=\langle0| E_{i}(\tau,\mathbf{x}_{\alpha}) E_{j}(\tau',\mathbf{x}_\beta) |0 \rangle$, $(i,j=\rho,\varphi,z$, and $\alpha,\beta=1,2)$ as 
\begin{eqnarray}\label{liangdianhanshu-1}
G^{(11)}_{zz}&=&G^{(22)}_{zz} \nonumber \\
&=&\frac{\gamma^2\{2R^2+\gamma^2 u^2+2R^4\Omega^2+R^2[(-2-2R^2\Omega^2+\gamma^2 u^2\Omega^2)\cos h-4h \sin h]\}}{\pi^2(-2R^2+\gamma^2 u^2+2R^2\cos h)^3}, \\
G^{(12)}_{zz}&=&G^{(21)}_{zz} \nonumber \\
&=&\frac{\gamma^2\{L^2-2R^2-\gamma^2 u^2-2R^4\Omega^2+R^2[2-(L^2-2R^2+\gamma^2 u^2)\Omega^2]\cos h+4hR^2\sin h\}}{\pi^2(L^2+2R^2-\gamma^2 u^2-2R^2\cos h)^3},\\
G^{(11)}_{\rho\rho}&=&G^{(22)}_{\rho\rho}\nonumber \\
&=&\frac{\gamma^2\{[u^2\gamma^2-2(R^2+R^4\Omega^2)]\cos h+R^2[2+(2R^2+u^2\gamma^2)\Omega^2-4h \sin h]\}}{\pi^2(-2R^2+u^2\gamma^2+2R^2\cos h)^3},\\
G^{(12)}_{\rho\rho}&=&G^{(21)}_{\rho\rho}\nonumber \\
&=&\frac{\gamma^2\{[-L^2-u^2\gamma^2+2(R^2+R^4\Omega^2)]\cos h+R^2[-2+(L^2-2R^2-u^2\gamma^2)\Omega^2+4h\sin h]\}}{\pi^2(L^2+2R^2-u^2\gamma^2-2R^2\cos h)^3},\nonumber\\\\
G^{(12)}_{\rho z}&=&-G^{(12)}_{z\rho}=-G^{(21)}_{\rho z}=G^{(21)}_{z\rho}
=\frac{4LR\gamma^2\sin\frac{h}{2}[h\cos\frac{h}{2}-(1+R^2\Omega^2)\sin\frac{h}{2}]
}{\pi^2(L^2+2R^2-u^2\gamma^2-2R^2\cos h)^3},\\
G^{(11)}_{\varphi\varphi}&=&G^{(22)}_{\varphi\varphi}=\frac{-2R^2+(2R^2+u^2\gamma^2)\cos h}{\pi^2(-2R^2+u^2\gamma^2+2R^2\cos h)^3},\\
G^{(12)}_{\varphi\varphi}&=&G^{(21)}_{\varphi\varphi}=\frac{-2R^2+(L^2+2R^2+u^2\gamma^2)\cos h}{\pi^2(-L^2-2R^2+u^2\gamma^2+2R^2\cos h)^3},\\
G^{(11)}_{\rho\varphi}&=&G^{(22)}_{\rho\varphi}=-G^{(11)}_{\varphi\rho}=-G^{(22)}_{\varphi\rho}
=\frac{u\gamma^2[2R^2\Omega(-1+\cos h)+u\gamma\sin h]}{\pi^2(-2R^2+u^2\gamma^2+2R^2\cos h)^3},\\
G^{(12)}_{\rho\varphi}&=&-G^{(12)}_{\varphi\rho}=G^{(21)}_{\rho\varphi}=-G^{(21)}_{\varphi\rho}
=\frac{-2\gamma R^2h(-1+\cos h)-\gamma(L^2+u^2\gamma^2)\sin h}{\pi^2(L^2+2R^2-u^2\gamma^2-2R^2\cos h)^3},\\
G^{(12)}_{z\varphi}&=&G^{(12)}_{\varphi z}=-G^{(21)}_{z\varphi}=-G^{(21)}_{\varphi z}
=\frac{2LR\gamma(h\cos h-\sin h)}{\pi^2(L^2+2R^2-u^2\gamma^2-2R^2\cos h)^3},\label{liangdianhanshu-2}
\end{eqnarray}
where we have defined $u=\tau-\tau'$, and $h=u\gamma\Omega$ for brevity.

For circularly accelerated atoms, the radius $R$ and the angular velocity $\Omega$ can be expressed with the velocity $v$ and acceleration $a$ as $R=\gamma^2v^2/a$, and $\Omega=v/R=a/\gamma^2v$. Plugging the above two equations and the Lorentz factor $\gamma=1/\sqrt{1-v^2}$ into Eqs. (\ref{liangdianhanshu-1})-(\ref{liangdianhanshu-2}), and taking the ultrarelativistic limit $v\rightarrow1$, the field correlation functions become
\bea\label{2p-limit-1}
G^{(11)}_{zz}&=&G^{(22)}_{zz}=\frac{24(72+6a^2u^2+a^4u^4)}{\pi^2u^4(12+a^2u^2)^3},  \\
G^{(12)}_{zz}&=&G^{(21)}_{zz}= \frac{24[-36L^2(2+a^2u^2)+u^2(72+6a^2u^2+a^4u^4)]}{\pi^2(-12L^2+12u^2+a^2u^4)^3}, \\
G^{(11)}_{\rho\rho}&=&G^{(22)}_{\rho\rho}=\frac{24(72-30a^2u^2+a^4u^4)}{\pi^2u^4(12+a^2u^2)^3},
\eea
\bea
G^{(12)}_{\rho\rho}&=&G^{(21)}_{\rho\rho}=\frac{24[-36L^2(-2+a^2u^2)+u^2(72-30a^2u^2+a^4u^4)]}{\pi^2(-12L^2+12u^2+a^2u^4)^3},\\
G^{(12)}_{\rho z}&=&-G^{(12)}_{z\rho}=-G^{(21)}_{\rho z}=G^{(21)}_{z\rho}=\frac{288aLu^2(-6+a^2u^2)}{\pi^2(-12L^2+12u^2+a^2u^4)^3},\\
G^{(11)}_{\varphi\varphi}&=&G^{(22)}_{\varphi\varphi}=\frac{144(12-5a^2u^2)}{\pi^2u^4(12+a^2u^2)^3},\\
G^{(12)}_{\varphi\varphi}&=&G^{(21)}_{\varphi\varphi}=\frac{144(12L^2+12u^2-5a^2u^4)}{\pi^2(-12L^2+12u^2+a^2u^4)^3},\\
G^{(11)}_{\rho\varphi}&=&G^{(22)}_{\rho\varphi}=-G^{(11)}_{\varphi\rho}=-G^{(22)}_{\varphi\rho}=\frac{144a(-12+a^2u^2)}{\pi^2u^3(12+a^2u^2)^3},\\
G^{(12)}_{\rho\varphi}&=&-G^{(12)}_{\varphi\rho}=G^{(21)}_{\rho\varphi}=-G^{(21)}_{\varphi\rho}=\frac{144au(12L^2+12u^2-a^2u^4)}{\pi^2(-12L^2+12u^2+a^2u^4)^3},\\
G^{(12)}_{z\varphi}&=&G^{(12)}_{\varphi z}=-G^{(21)}_{z\varphi}=-G^{(21)}_{\varphi z}=\frac{1152La^2u^3}{\pi^2(-12L^2+12u^2+a^2u^4)^3},
\label{2p-limit-2}
\eea
where $u=\tau-\tau'$.




\section{Master equation  in the coupled basis}\label{app2}

To investigate the dynamics of the two-atom system, we work in the coupled basis $\{|G\rangle=|00\rangle,|A\rangle=\frac{1}{\sqrt{2}}(|10\rangle-|01\rangle),|S\rangle=\frac{1}{\sqrt{2}}(|10\rangle+|01\rangle),|E\rangle=|11\rangle\}$, and then a set of equations which are decoupled from other matrix elements can be derived from Eq. (\ref{zhufangchen}) as \cite{ZF2002}
\begin{eqnarray}
\label{rgg}
\dot{\rho}_{GG}&=&-2(A_1+A_2-B_1-B_2)\rho_{GG}+(A_1+A_2-A_3-A_4+B_1+B_2-B_3-B_4)\rho_{AA}\nonumber\\
&&+(A_1+A_2+A_3+A_4+B_1+B_2+B_3+B_4)\rho_{SS}+(A_1-A_2-A_3+A_4+B_1\nonumber\\
&&-B_2-B_3+B_4)\rho_{AS}+(A_1-A_2+A_3-A_4+B_1-B_2+B_3-B_4)\rho_{SA},\\
\dot{\rho}_{EE}&=&-2(A_1+A_2+B_1+B_2)\rho_{EE}+(A_1+A_2-A_3-A_4-B_1-B_2+B_3+B_4)\rho_{AA}\nonumber\\
&&+(A_1+A_2+A_3+A_4-B_1-B_2-B_3-B_4)\rho_{SS}+(-A_1+A_2+A_3-A_4+B_1\nonumber\\
&&-B_2-B_3+B_4)\rho_{AS}+(-A_1+A_2-A_3+A_4+B_1-B_2+B_3-B_4)\rho_{SA},\\
\dot{\rho}_{AA}&=&-2(A_1+A_2-A_3-A_4)\rho_{AA}+(A_1+A_2-A_3-A_4-B_1-B_2+B_3+B_4)\rho_{GG}\nonumber\\
&&+(A_1+A_2-A_3-A_4+B_1+B_2-B_3-B_4)\rho_{EE}+(-B_1+B_2+B_3-B_4)\rho_{AS}
\nonumber\\
&&+(-B_1+B_2-B_3+B_4)\rho_{SA},\label{raa}
\eea
\bea
\dot{\rho}_{SS}&=&-2(A_1+A_2+A_3+A_4)\rho_{SS}+(A_1+A_2+A_3+A_4-B_1-B_2-B_3-B_4)\rho_{GG}\nonumber\\
&&+(A_1+A_2+A_3+A_4+B_1+B_2+B_3+B_4)\rho_{EE}+(-B_1+B_2+B_3-B_4)\rho_{AS}\nonumber\\
&&+(-B_1+B_2-B_3+B_4)\rho_{SA},\label{rss}\\
\dot{\rho}_{AS}&=&(A_1-A_2-A_3+A_4-B_1+B_2+B_3-B_4)\rho_{GG}+(-A_1+A_2+A_3-A_4-B_1\nonumber\\
&&+B_2+B_3-B_4)\rho_{EE}+(-B_1+B_2-B_3+B_4)(\rho_{AA}+\rho_{SS})-2(A_1+A_2)\rho_{AS},\\
\dot{\rho}_{SA}&=&(A_1-A_2+A_3-A_4-B_1+B_2-B_3+B_4)\rho_{GG}+(-A_1+A_2-A_3+A_4-B_1\nonumber\\
&&+B_2-B_3+B_4)\rho_{EE}+(-B_1+B_2+B_3-B_4)(\rho_{AA}+\rho_{SS})-2(A_1+A_2)\rho_{SA},\nonumber\\
\dot{\rho}_{GE}&=&-2(A_1+A_2)\rho_{GE},\qquad\qquad \dot{\rho}_{EG}=-2(A_1+A_2)\rho_{EG},
\label{reg}
\end{eqnarray}

where $\rho_{IJ}=\langle I|\rho|J\rangle,I,J\in\{G,E,A,S\}$.

\end{document}